\documentclass[letterpaper, 10 pt, conference]{ieeeconf} 

\IEEEoverridecommandlockouts                              
\usepackage{bm}
\usepackage{amsmath}
\usepackage{accents}
\usepackage{amssymb,amsfonts}
\usepackage{graphicx}
\usepackage{textcomp}
\usepackage{xcolor}
\usepackage{optidef}
\usepackage{hyperref}
\usepackage{cleveref}
\usepackage[linesnumbered]{algorithm2e}
\usepackage[textsize=tiny]{todonotes}
\usepackage{subcaption}
\newcommand{\ubar}[1]{\text{\b{$#1$}}}

\newcommand{\Roa}{\tilde{\mathcal{R}}_t}

\newcommand{\Rvec}{\tilde{\mathbb R}}
\newcommand{\tcurrent}{t_{\text{cur}}}
\newcommand{\tstart}{t_{\text{start}}}
\newcommand{\test}{t_{\text{est}}}
\newcommand{\bsteps}{b_{\text{steps}}}

\newtheorem{theorem}{Theorem}[section]
\newtheorem{lemma}[theorem]{Lemma}

\newtheorem{corollary}[theorem]{Corollary}
\newtheorem{definition}[theorem]{Definition}

\def\underacc #1/#2{\mathchoice{\uacc\textstyle{#1}{#2}}{\uacc\textstyle{#1}{#2}}
                     {\uacc\scriptstyle{#1}{#2}}{\uacc\scriptscriptstyle{#1}{#2}}}
\def\uacc#1#2#3{\mathop{#2{}}\limits_{#1#3{}}}

\title{\LARGE \bf
TTT: A Temporal Refinement Heuristic for Tenuously Tractable Discrete Time Reachability Problems
}

\author{Chelsea Sidrane$^{1,2}$, 
Jana Tumova$^{1,2}$
\thanks{C. Sidrane and J. Tumova are with the $^1$Division of Robotics, Perception and Learning, Intelligent Systems Department, School of Electrical Engineering \& Computer Science, and $^2$\href{https://www.digitalfutures.kth.se/}{Digital Futures} at
        KTH Royal Institute of Technology, Brinellvägen 8, 114 28 Stockholm, Sweden
        {\tt\small chelse@kth.se, tumova@kth.se}}
}%

\begin{document}

\def\BibTeX{{\rm B\kern-.05em{\sc i\kern-.025em b}\kern-.08em
    T\kern-.1667em\lower.7ex\hbox{E}\kern-.125emX}}
\markboth{\journalname, VOL. XX, NO. XX, XXXX 2017}
{Author \MakeLowercase{\textit{et al.}}: Preparation of Papers for IEEE Control Systems Letters (August 2022)}

\maketitle
\thispagestyle{empty}
\pagestyle{empty}

\begin{abstract}

Reachable set computation is an important tool for analyzing control systems.
Simulating a control system can show general trends, but a formal tool like reachability analysis can provide guarantees of correctness. 
Reachability analysis for complex control systems, e.g., with nonlinear dynamics and/or a neural network controller, is often either slow or overly conservative.
To address these challenges,  much literature has focused on spatial  refinement, i.e.,  tuning the discretization of the input sets and intermediate reachable sets.
This paper introduces the idea of temporal refinement: automatically choosing \emph{when} along the horizon of the reachability problem to execute slow symbolic queries which incur less approximation error versus fast concrete queries which incur more approximation error.
Temporal refinement can be combined with other refinement approaches as an additional tool to trade off tractability and tightness in approximate reachable set computation.
We introduce a temporal refinement algorithm and demonstrate its effectiveness at computing approximate reachable sets for nonlinear systems with neural network controllers.
 We calculate reachable sets with varying computational budget and show that our algorithm can generate approximate reachable sets with a similar amount of error to the baseline in 20-70\% less time.

\end{abstract}


\section{Introduction}
 Every controller needs to be analyzed for performance and reliability. 
Simulation is useful to get an idea of the general behavior of a control policy,  but only formal  analysis can guarantee correctness with respect to a model of the system and a specification.
In this paper, we focus on reach-avoid specifications which dictate a goal set the system must reach and/or an avoid set it must \textit{not} reach.
 For a given discrete time dynamical system with a control policy
 and set of  possible starting states 
the forward reachable set at time $t$ is the set of states that the system could reach $t$ steps into the future.
Given a final set instead of an initial set, the $t$-step backward reachable set is the set of states that will reach the final set after $t$ steps.
For simple settings like linear dynamical systems 
and  convex polytopic initial sets 
reachability analysis is straightforward and fast~\cite{frehse2011spaceex}.
However, many modern control systems operate on nonlinear systems~\cite{slotine1991applied}, and increasingly contain neural network components. 
Reachability methods for such complex systems may not exist, and if they do, they almost always require under or over-approximation (also referred to as inner and outer approximation) of the reachable set as the exact set is not computable. 
It is desirable to produce tight reachable sets which have as little approximation error as possible,  but this often trades off with tractability --   the computation time needed to compute the reachable set.

\begin{figure}[t]
    \centering
    \includegraphics[width=.99\columnwidth,trim=.5cm 1.2cm 3cm 1.5cm,clip]{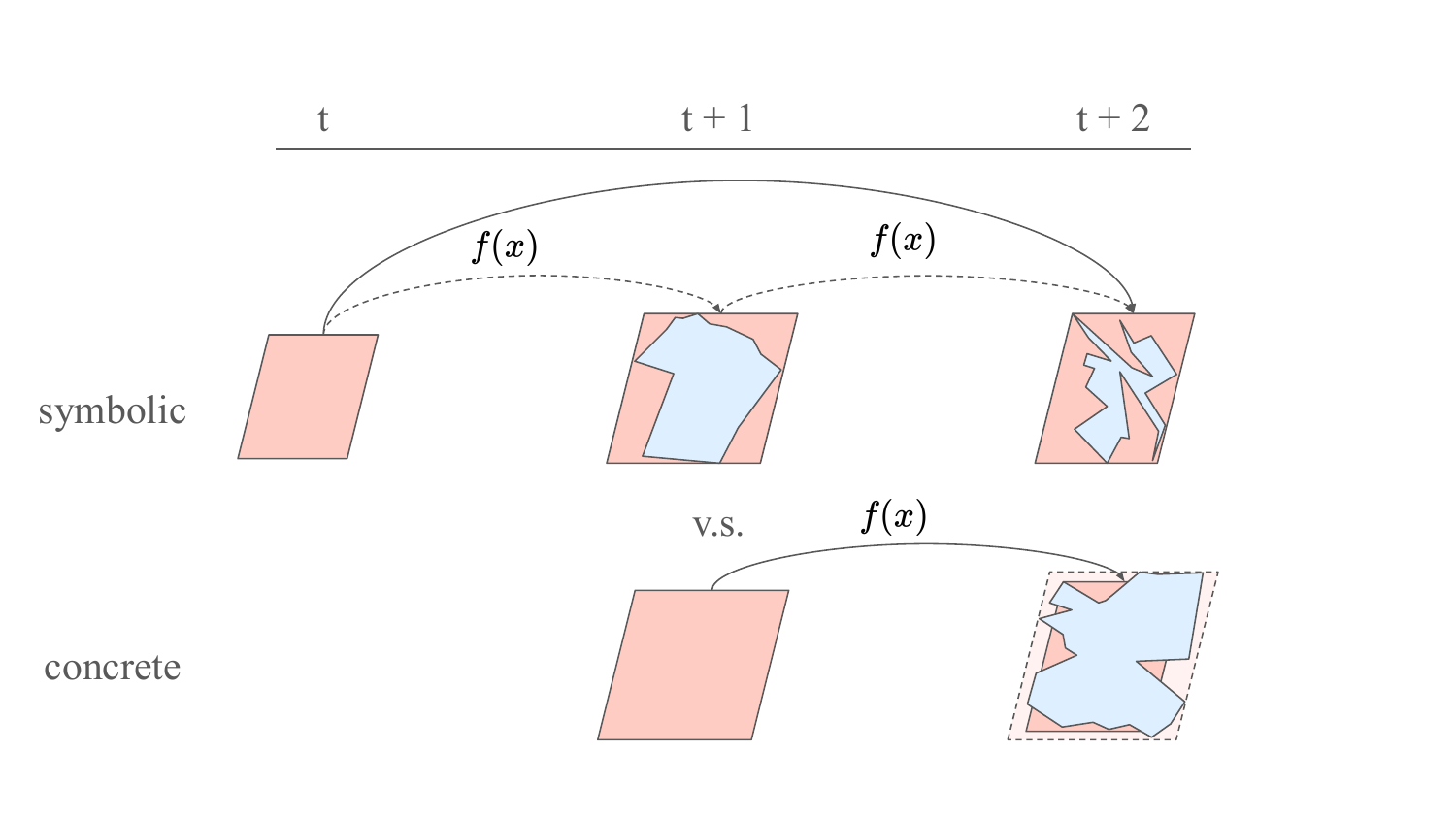}
    \caption{\footnotesize An illustration of overapproximate forward reachability. 
    A symbolic query uses multiple copies of the dynamical system update function $f(x)$ within a single computation to 
    produce tighter final sets. 
    This concept is central to the refinement algorithm.}
    \label{fig:concept}
\end{figure}

There have been many approaches in the reachability literature to address tractability and tightness. 
One of the central concepts used is domain refinement~\cite{bansal2017hamilton, everett2023drip}.
Domain refinement refers to splitting the domain 
into subsets, e.g., through gridding, 
and computing the reachable set for each subset.  
Without domain refinement, the reachable set approximation would be unusably loose, but if gridded too finely the problem can become intractable as there is then exponential complexity in the number of state dimensions.

Another concept that has been used in the reachability literature is decomposition~\cite{chen2018decomposition, mitchell2003overapproximating, chen2016decomposed}.
Decomposition refers to re-writing the system dynamics 
as several uncoupled or loosely coupled subsystems.
If using gridding to produce tight set approximations, fine gridding can then be used on each low dimensional subsystem without the exponential cost of gridding the original high dimensional system. 

Lastly, many methods make use of pre-specified template sets in order to speed up computation~\cite{sidrane2022overt, kurzhanski2000ellipsoidal, kuhn1998rigorously, althoff2011zonotope}. 
In these methods, the tightest circumscribing (or inscribing) set of a particular type such as a hyperrectangle, ellipsoid, or zonotope is computed rather than computing the arbitrary non-convex true reachable set. 

In this work we make use of a concept 
known as symbolic reachability~\cite{sidrane2022overt} to define an algorithm for \textit{temporal refinement}.
Many discrete-time verification methods use \textit{ concrete reachability} which involves computing the approximate reachable set at time $t+1$ using only the approximate reachable set at time $t$ (or $t-1$ from $t$ in the backward case) and a single copy of the dynamical system update function~\cite{siefert2023successor}.
While  concrete reachability is  fast, it accumulates large approximation error quickly, a phenomenon known as the ``wrapping effect''~\cite{neumaier1993wrapping}. 
\textit{Symbolic reachability} can be used to mitigate this effect~\cite{sidrane2022overt}. 
Using symbolic reachability, the approximate reachable set at e.g., time $t+2$ is computed from the set at time $t$ using e.g., two copies of the dynamical system update function (see \cref{fig:concept}). 
This results in tighter sets than if the set at $t+2$ was computed using only the approximate set at time $t+1$ and a single copy of the update function.
It is possible to perform symbolic reachability for limited time horizons (temporal depths), each iteration computing a set farther into the future (or further in the past), 
but the problem will quickly approach intractability.
To address this tradeoff between tightness and tractability, the authors of \cite{sidrane2022overt} perform hybrid-symbolic reachability which uses mixtures of  concrete and symbolic computations in a hand-tuned, ad-hoc manner. 

In this paper, we contribute a \textit{temporal refinement} algorithm to automate hybrid-symbolic reachability analysis for tenuously tractable discrete-time dynamical systems.
The  temporal depth and end timestep of each query are automatically selected to produce tight reachable sets given the approximate remaining computational budget and approximate projected total computation length.
Our algorithm is an orthogonal contribution to existing strategies for balancing  tightness  and tractability because it may be combined with concepts such as domain refinement, system decomposition, and pre-specified template sets; among others.
\footnote{Note that while temporal refinement may sound similar to timestep adaptation  for continuous systems~\cite{wetzlinger2022adaptive}, it is not closely related.
In a continuous time system,  the length of time step $\Delta t$  is an algorithm hyperparameter that maybe adapted  according to error requirements, but in a discrete time system, there is either no specific step length, only time indices, $t=0,t=1, \ldots$ or there is a fixed timestep length $\Delta t$. 
}

We demonstrate our algorithm for temporal refinement on overapproximate forward reachability analysis, but it is similarly applicable to backward reachability analysis and underapproximation.
 Specifically, we treat a challenging class of problems:  computing forward reachable sets for discrete time nonlinear  dynamical systems with neural network control policies; also called Neural Feedback Loops (NFLs).
For this class of problem, there has been work on computing forward reachable sets~\cite{sidrane2022overt}, backward reachable sets~\cite{rober2023backward}, performing domain refinement~\cite{rober2023hybrid, everett2023drip, xiang2020reachable}, and using exotic template polyhedron~\cite{manzanas2021verification} but there remains difficulty balancing tractability and tightness making it a good class of problem on which to demonstrate our technique.
We 
use several examples from the literature and show that our approach can produce a range of overapproximation error values depending on the allotted approximate computational budget, including computing reachable sets 20-70\% faster than previous hand-tuned approaches for the same amount of error in the reachable set approximation.

\section{Problem Definition}
Consider a discrete time dynamical system with state $\vec{x} \in R^d$ governed by update equation $\vec{x}_{t+1} = f(\vec{x}_{t}, \vec{u}_t)$ with $\vec{u} \in R^m$. In this paper, we focus on nonlinear update functions $f$ but linear systems may also be considered. 
The control input $\vec{u}_t = c(\vec{x}_t)$ comes from a feedback control policy.

We seek to compute the forward reachable sets $\mathcal R_t$ of the controlled system at future timesteps $t \in 1, 2, 3, \ldots$. 
The forward reachable set at time $t_n$, $n$ steps into the future from time $0$ is defined as follows:
\begin{align}
&\mathcal R_n \triangleq \{ \vec{x}_n \mid \vec{x}_n = f(\vec{x}_{n-1}, c(\vec{x}_{n-1})), \\ \nonumber
&\vec{x}_{n-1} = f(\vec{x}_{n-2}, c(\vec{x}_{n-2})),\\ \nonumber
& \ldots, \\ \nonumber
&\vec{x}_{1} = f(\vec{x}_{0}, c(\vec{x}_{0})), \\ \nonumber
&\vec{x}_0 \in \mathcal X_0\} \nonumber
\end{align}
where $\mathcal X_0$ is the initial set of states.
As aforementioned, methods exist to compute the exact reachable set $\mathcal R_t$ for classes of linear or piecewise linear systems, but for
arbitrary smooth nonlinear dynamics, approximation must be used.
This approximation may be overapproximation $\tilde{\mathcal{R}}_t \supseteq \mathcal R_t$, or underapproximation $\underacc \mathcal{R}_t/\tilde \subseteq \mathcal{R}_t$. 

The overapproximation error $e$ at a given time $t$ is defined 
\begin{equation} \label{eq:oaedef}
    e = \frac{m(\tilde{\mathcal{R}}_t)}{m(\mathcal R_t)}
\end{equation}
where $m$  is an  error metric and the desired value is $e=1$.
The error over a finite time horizon $t\in\{1\ldots n\}$ is defined
\begin{equation} \label{eq:total_err}
    {e} = \frac{\sum_{t=1}^n m(\tilde{\mathcal{R}}_t)}{\sum_{t=1}^n m({\mathcal{R}_t})}
\end{equation}
Thus, the problem may be stated as follows. \vspace{5pt} \\
\textit{Problem 1:}
\textit{Compute a vector of $n$ approximate reachable sets $\tilde{\mathbb R}~\triangleq~[\tilde{\mathcal{R}}_1, \ldots, \tilde{\mathcal{R}}_n]$ using hybrid-symbolic reachability analysis with an automatically chosen mixture of concrete and symbolic queries so as to minimize approximation error (\cref{eq:total_err}) while approximately adhering to a pre-specified computational budget $b$.}


\section{Refinement Algorithm}
\begin{figure}[ht!]
    \centering
    \includegraphics[width=0.85\columnwidth,trim=6cm .7cm 6cm .5cm,clip]{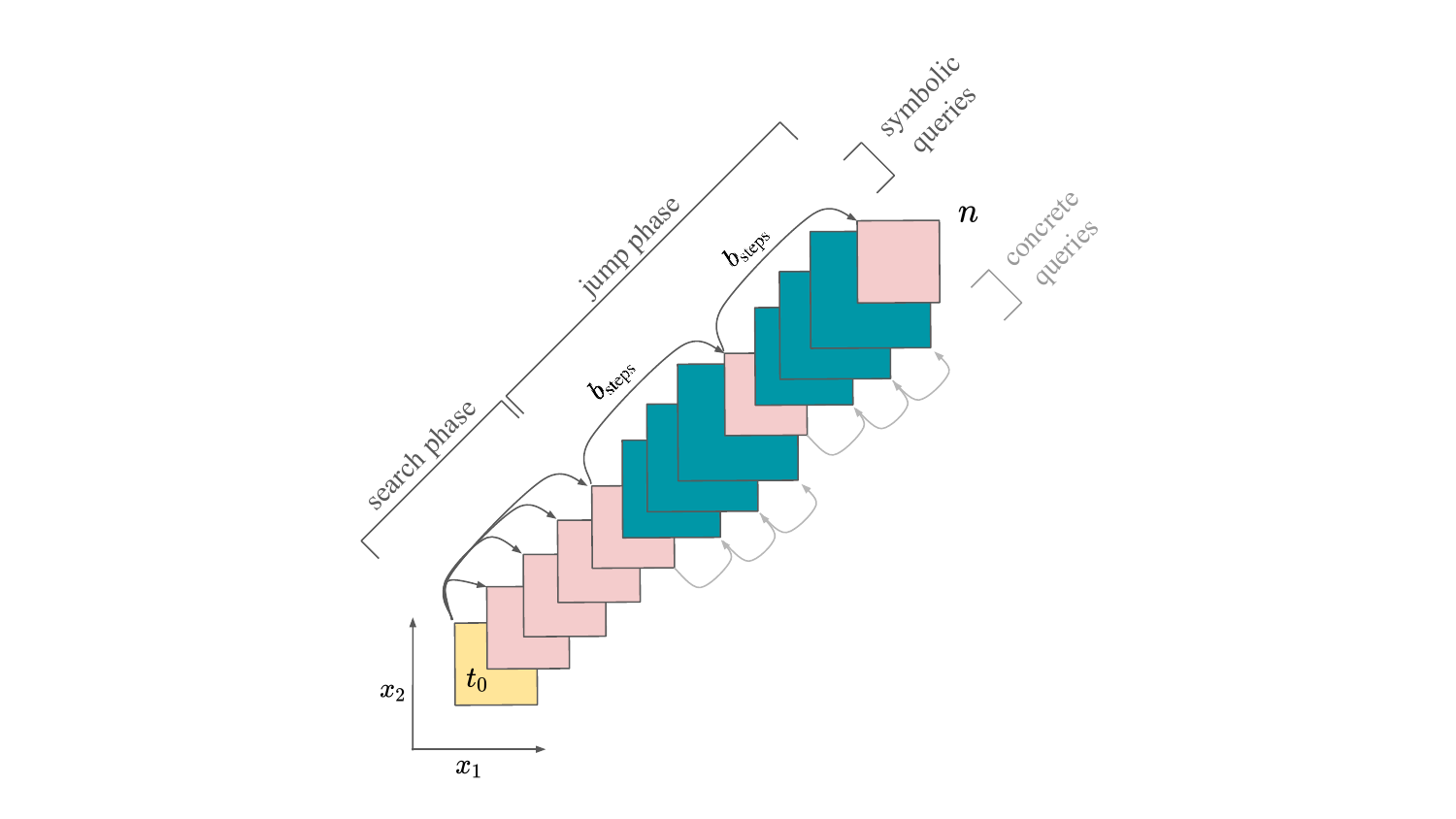}
        \caption{\footnotesize An illustration of Alg.~\ref{alg:overall}. In this example $b_{\text{steps}}=4$.} \label{fig:initial_reach}
\end{figure}

We introduce Alg.~\ref{alg:overall}, illustrated in \cref{fig:initial_reach}, as a solution to Problem 1.
Alg.~\ref{alg:overall} automates hybrid symbolic reachability analysis by automatically selecting points $t_{i} \in \{1,\ldots,n\}$ at which to perform symbolic reachability queries, and temporal depths $\bsteps \in \{1\ldots t_i\}$ for those symbolic queries so as to minimize approximation error. 
At a high level, the algorithm begins in the ``search'' phase and searches for an estimate of the longest tractable temporal depth (symbolic horizon) $b_\text{steps}$.
The algorithm then moves to a ``jump'' phase and performs symbolic queries of that temporal depth or shorter until the set at the desired final time $n$ has been reached. 
More detail is provided in Alg.~\ref{alg:overall}, where $\mathcal X_0$ is the initial set for reachability analysis, $b$ is the approximate time budget in seconds, symbolic\_reach is a chosen symbolic reachability procedure, and $q$ is a query object consisting of the transition system and reachability horizon. 
For our chosen class of example problems, $q$ consists of a nonlinear dynamics function $f$, a neural network controller $c(x) := NN(x)$, a final time horizon $n$, a variable to store current temporal depth $h$, and hyperparameters, e.g., controlling tightness of the nonlinear overapproximation. 
The method returns $\tilde{\mathbb R}$ a vector of approximate reachable sets $\Roa, \forall t \in [1\ldots n]$.

The motivation for the structure of the algorithm comes partially from the initial work on hybrid-symbolic reachability~\cite{sidrane2022overt} which used long symbolic jumps with one-step concrete queries to generate the sets in between. 
Long symbolic jumps produce tight approximations of the reachable set and beginning symbolic jumps from tight approximations limits the wrapping effect. 
However, in this paper, a reasonable symbolic ``jump'' size is not known \textit{a priori} so we first search for a long jump size. 
The search procedure also has the benefit of limiting reachable set growth.
Small errors early in the time horizon compound over many timesteps, and because the search phase produces tight reachable sets for early steps, it limits error compounding. 


Alg.~\ref{alg:overall} begins in the ``search'' phase  with a temporal depth of $\bsteps = 1$ (line 3) beginning from the set $\mathcal X_0$ at time $t_0$ (lines 4-5).
A symbolic reachability query is then performed (lines 7-8), and  the new set(s) are then logged (lines 9-13).
If there is enough time budget  to continue searching, $\bsteps$ will increment by $1$ (line 16).
Symbolic reachable set computation queries are performed from the initial set $\mathcal X_0$ at time $t_0$ to increasingly large temporal depths / numbers of steps $\bsteps$ into the future until it is estimated that trying a temporal depth one step longer would put the entire reachable set procedure over budget.
Once this occurs,  the algorithm will enter the ``jump'' phase and perform symbolic queries beginning from the set at $\tstart := t_{\text{cur}}$ (line 18-19) of temporal depth $\bsteps$ or shorter until the set at the desired final time $n$ has been reached (line 6). 
 Line 21 keeps $\bsteps$ within correct range; and line 22 updates the early stopping timeout.
 The early stopping timeout limits budget overruns and is hand-tuned. 
 The algorithm returns reachable sets from $t=1\ldots n$.

 \begin{algorithm}
\DontPrintSemicolon
\caption{Reachability for NFLs with automatic temporal refinement\label{alg:overall}}
\SetKwFunction{vf}{RefinedReach}
\SetKwProg{Fn}{function}{:}{}
\SetKwComment{Comment}{\#}{}
\Fn{\vf{$q$, $\mathcal X_0$, $b$, symbolic\_reach}}{

set\_initial\_timeout!($b$, $q$) \\
phase $\gets$ ``search''; $b_\text{steps} \gets 1$ \\
$t_\text{start} \gets$ 0; $t_\text{cur} \gets$ 0;   \\
$\mathcal X_\text{start} \gets \mathcal X_0$; $\Rvec \gets $ [] \\
\While{$t_{\text{cur}} < n$}{
    q.h $\gets$ $b_\text{steps}$\\
    data, $\tilde{\mathbb{R}}_{\text{out}}$ $\gets$ symbolic\_reach($q, \mathcal X_\text{start}$)\\
    \uIf{phase == ``search''}{
        push!($\Rvec$, $\tilde{\mathbb{R}}_{\text{out}}[end]$) 
    }
    
    \Else{
        push!($\Rvec$, $\tilde{\mathbb{R}}_{\text{out}}$)\\
    }
    $t_\text{cur} \gets t_\text{start} + b_\text{steps}$ \\
    $b \gets b~-~$elapsed()\\
    $b_{\text{steps}_0}$, phase $\gets$ calc\_steps($t_\text{start}, b_\text{steps}, b$, data, $\test$, $n$, phase, status(q))\\
    \If{phase == ``jump''}{
        $t_\text{start} \gets t_\text{start} + b_\text{steps}$\\
        $\mathcal X_\text{start} \gets \Rvec[t_\text{start}$]\\
    }
    $b_\text{steps} \gets \min(b_{\text{steps}_0}, n-t_\text{start})$\\
        update\_timeout!($q$, $b$)\\
    } 
return $\tilde{\mathbb R}$
\;}
\end{algorithm}

One of the key pieces of Alg.~\ref{alg:overall} is the subroutine calc\_steps (line 16) which determines whether to continue  searching for a longer temporal depth $b_\text{steps}$ or to move to the ``jump'' phase. 
The complete algorithm for calc\_steps is included in the Appendix (alg.~\ref{alg:calc_qsteps}) and described here.
calc\_steps first estimates the time needed to compute one symbolic step, $\test$, assuming that the time to compute a symbolic query of $m$  steps is roughly $\test*m$.
This is a simplification that is not strictly accurate -- the time a symbolic query takes may be in the worst case exponential in temporal depth, and timing depends on the particular input set --  but it leads to roughly accurate budgeting when $\test$ is updated at each iteration. 
 The algorithm then estimates the time that future queries will take and uses this to determine if there is enough time budget to continue searching.
This time estimate includes a symbolic query from $t_0$ to $\bsteps + 1$ plus the time to finish finish the temporal horizon in the ``jump'' phase.
If the remaining budget $b$ is estimated to be large enough, the algorithm increments $\bsteps \gets \bsteps + 1$ and if not, the algorithm enters the ``jump'' phase.
If the solver stops early, the phase is also switched to ``jump''.
 Early stopping is used to ensure the budget is approximately respected given that the estimated time per symbolic step $\test$ is not an upper bound. 
Note that  long symbolic queries generally  produce tight reachable sets but shorter queries can produce tighter sets if the starting set of the shorter query is tighter itself. 


%
\section{Overapproximate Forward Reachability}
 Our refinement algorithm is applicable to both forwards and backwards reachability and both over (outer) and under (inner)  approximation, but in this paper we demonstrate our refinement algorithm on overapproximate forward reachability of nonlinear neural feedback loops (NFLs).
 In particular, we use the reachability approach presented in \cite{sidrane2022overt}.
 To briefly summarize, the approach rewrites high dimensional nonlinear dynamics into many one dimensional functions. 
 It then computes  tight piecewise linear bounds for each smooth nonlinear 1d function and then encodes these bounds as constraints in a mixed integer linear program (MILP).
 The neural network control policy is limited to ReLU activation functions making it a piecewise linear function that is then encoded as constraints in the MILP as well.
A hyperrectangular template set is then optimized  to obtain an overapproximation of the reachable set.
The hyperrectangular set is constructed by finding the minimum and maximum value of each state variable $\vec{x}^{(i)}_t, i\in 1\ldots d$.
The mixture of symbolic and concrete queries is determined by the refinement algorithm presented here, but the procedure to approximate nonlinear functions and compute approximate reachable sets using MILPs comes from \cite{sidrane2022overt} where further details can be found.

\section{Soundness, Completeness, Complexity}
The section serves to demonstrate the soundness, completeness in the sense of termination, and complexity of the proposed Algorithm~\ref{alg:overall} for solving Problem 1. 
We first show two supporting lemmas, \ref{lem:oa} and \ref{lem:stopping}, before asserting the main theorem \ref{the:main}.
\lemma{Alg.~\ref{alg:overall} produces sets $\tilde{\mathcal{R}}_t \in \tilde{\mathbb{R}}$ which overapproximate the true reachable set $\mathcal R_t \subseteq \tilde{\mathcal{R}}_t$ at each time $t \in \{1,\ldots,n\}$.\\ 
\proof{This follows directly from the overapproximation property of the symbolic\_reach procedure being called. E.g., in the examples presented here, from \cite{sidrane2022overt}. $\square$}\\
Note that while Alg.~\ref{alg:overall} is written for overapproximate sets, it could be easily reformulated to compute underapproximations $\underacc \mathcal{R}_t/\tilde \subseteq \mathcal{R}_t$ and an analogous lemma would hold. \label{lem:oa}}
\lemma{Alg.\ref{alg:overall} will terminate producing $n$ reachable sets $\tilde{\mathbb{R}} \triangleq [\tilde{\mathcal{R}}_1,\ldots,\tilde{\mathcal{R}}_n] $
under the assumption that the true reachable set is non-emtpy.\label{lem:stopping}}\\
\proof{
All input variables are bounded and therefore each MIP is guaranteed to terminate in finite time when early stopping is not used.
When early stopping is used, the algorithm will extend the compute time until a finite feasible solution with $<50\%$ relative duality gap has been found.
Such a solution is guaranteed to exist and to be found in finite time. $\square$
}
 \theorem{Alg. 1 performs hybrid-symbolic reachability analysis with an automatically chosen mixture of concrete and  symbolic queries to produce $n$ overapproximate reachable sets $\tilde{\mathbb{R}} \triangleq [\tilde{\mathcal{R}}_1,\ldots,\tilde{\mathcal{R}}_n]$.\label{the:main}}\\
 \proof{See lemmas 5.1 and 5.2. $\square$} \\
We note that in addition to producing $n$ overapproximate reachable sets, Alg.~\ref{alg:overall} also heuristically minimizes overapproximation error and heuristically adheres to the budget.

Next, we reason about the behavior of Alg.~\ref{alg:overall} in the limit of infinite budget in Corollary~\ref{col:blim} and the worst-case complexity of Alg.~\ref{alg:overall} in Corollary~\ref{col:complex}.
 \definition{The tightest sequence of approximate reachable sets $\tilde{\mathbb{R}}|$ possible given a fixed, separable~\cite{chen2023one} subroutine symbolic\_reach is the sequence computed with pure symbolic reachability wherein each query begins from $t=0,~\mathcal{X}_0$ and extends to the current time $t$ for $t \in \{1,\ldots,n\}$.
\corollary{In the limit when the time budget $b\rightarrow \infty$, the algorithm will produce $\tilde{\mathbb{R}}|$.\label{col:blim}}\\
\proof{As the time budget $b \rightarrow \infty$, Alg.~\ref{alg:overall} will stay in the search phase until sets for the entire horizon $n$ have been computed. In the search phase, the algorithm performs pure symbolic reachability beginning from the initial set $\mathcal{X}_0 $. $\square$
}
\corollary{In the worst case Alg.~\ref{alg:overall} makes $O(n)$ calls to symbolic\_reach with worst-case temporal depth $O(n)$ where $n$ is the final time horizon.\label{col:complex}}\\
\proof{If we assume Alg.~\ref{alg:overall} finishes the search phase and chooses $\bsteps= b^*$ the algorithm will have made $b^*$ calls to symbolic\_reach, each of increasing length, i.e., $1,2,3,\ldots ,b^*$. 
In the jump phase the algorithm will then make $\frac{(s_{\text{left}})}{b^*} = \frac{(n-b^*)}{b^*} \approx \frac{n}{b^*}$ more calls to symbolic\_reach. 
This is in total for search and jump phases $b^* + \frac{n}{b^*}$ calls. 
In the worst case, $b^* \rightarrow n$ leading to $O(n)$ calls to symbolic\_reach where the longest and final call has temporal depth $n$ giving $O(n)$. $\square$} 

Note, however, that the complexity of reachability is strongly affected by the symbolic\_reach method. E.g., if symbolic\_reach solves a MILP with $m$ decision variables for $n=1$, $2m$ for $n=2$, etc. using branch and bound, the total complexity would be $\approx O(2^m + 2^{2m} + \cdots + 2^{nm}) \approx O(2^{nm})$. 
The average case complexity is in practice more reasonable  as is demonstrated in the numerical experiments section.

\section{Numerical Experiments}
Numerical experiments were run on examples from ~\cite{sidrane2022overt, tranarch} to assess the tightness and computational speed of the temporal refinement algorithm.
Four problems were used: the 
pendulum dynamical system with a controller of 2 layers, 25 neurons per layer (S1); the TORA dynamical system with a controller of 3 layers, 25 neurons per layer (T1); 
the car dynamical system with a controller of 1 layer with 100 neurons (C1) and a controller of 1 layer with 200 neurons (C2).
The reader is referred to \cite{sidrane2022overt} for the details on the neural network controlled dynamical systems.
 All experiments were run on a 10-core Apple M2 Pro notebook processor. 

\subsection{Results}
The temporal refinement algorithm was run  multiple times to assess how changing the time budget changes the approximate reachable set error.
\Cref{fig:err_vol} measures the  approximation error of the sets computed by the optimizer using the ratio of the sum of volumes:
\begin{equation}
    e_{\text{total}} = \frac{\sum_{t=1}^n \text{vol}(\Roa)}{\sum_{t=1}^n \text{vol}(\ubar{\mathcal{R}_t})}
    \label{eq:err_v}
\end{equation}
The true reachable set cannot be computed so error is computed with respect to $\ubar{\mathcal{R}}_t$ which is the hyperrectangular hull of 1,000,000 samples of the reachable set.
In \cref{fig:err_vol}, one can observe that the approximate error generally decreases with increasing budget.
Note that the trend is not monotonic because the algorithm is a heuristic.

We compare our algorithm to 
the hand-tuned hybrid symbolic approach used in \cite{sidrane2022overt} 
which is represented in \cref{fig:err_vol} as dotted lines.
One might expect that the hand-tuned approach would produce approximate reachable sets more quickly for a given amount of error.
In the hand-tuned approach, offline data was used to identify good symbolic intervals that must instead be identified at runtime with our approach.
However, it turns out that our heuristic can actually produce approximate reachable sets with similar amounts of error as the hand-tuned approach in 20-70\% less time, as shown in \cref{fig:err_vol}.
Furthermore, our approach is able to produce reachable sets for some problems such as the pendulum problem S1 that have 40\% less error than the hand-tuned approach when allowed to compute for longer than the time taken by the hand-tuned approach. 

\begin{figure}[h!]
    \centering
    \includegraphics[width=0.9\columnwidth,trim=0cm 0cm 0cm -.3cm,clip]{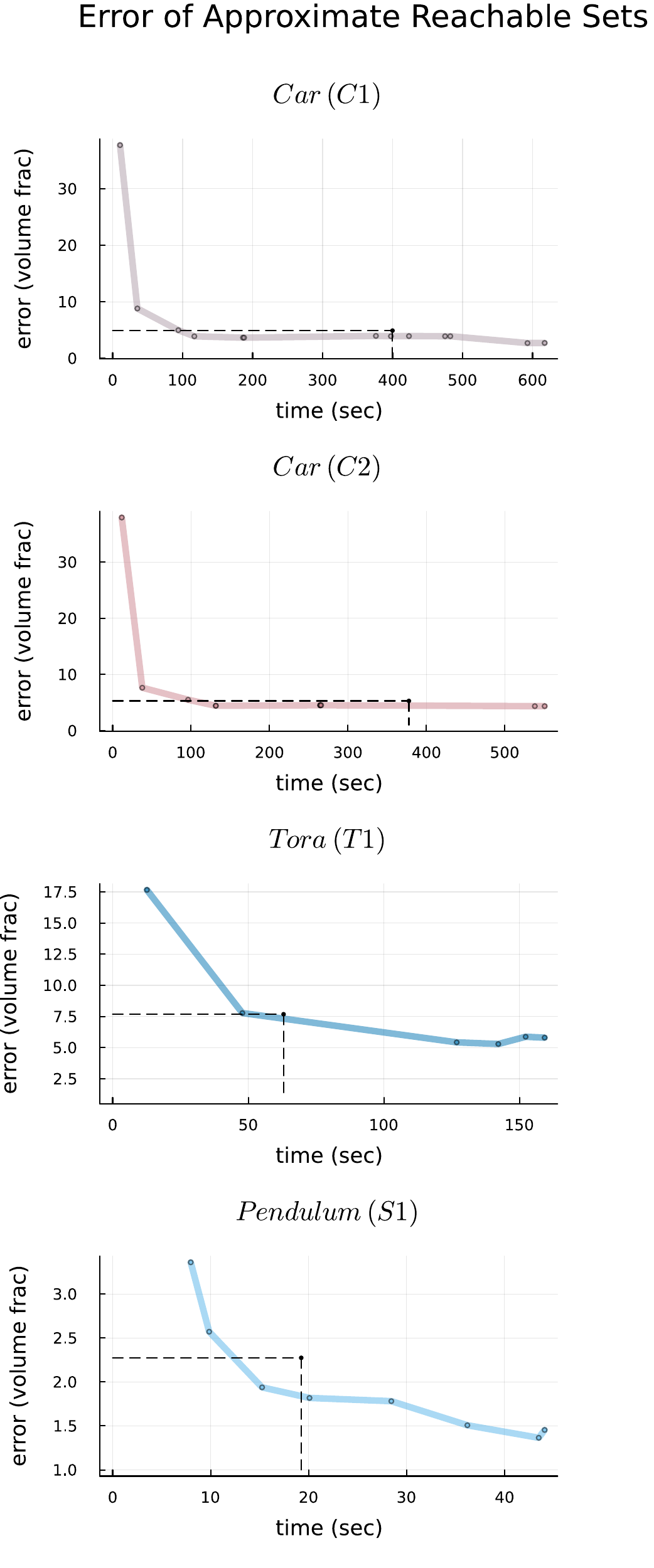}
\caption{\footnotesize Approximate reachable sets of various fidelity are produced depending on allotted time and are competitive with hand-tuned baseline \cite{sidrane2022overt} indicated with dashed lines.}
    \label{fig:err_vol}
\end{figure}

We also compare sets produced by our algorithm to concrete one step reachable sets. 
  Using one step sets is fast but leads to large approximation error.
Examples of the reachable sets computed by our algorithm and baselines can be seen in ~\cref{fig:resultsets} and \cref{fig:resultsets2}.
Set are compared both using \cref{eq:err_v} as well as the ratio of the sum of radii:
 \begin{equation}
      e_{\text{total}} = \frac{\sum_{t=1}^n \big(\sum_{i=1}^d\text{radius}_i(\Roa)\big)}{\sum_{t=1}^n \big(\sum_{i=1}^d\text{radius}_i(\ubar{\mathcal{R}_t} )\big)}
      \label{eq:err_r}
 \end{equation}
One step concrete sets are labeled 'naive' and one can observe that every set produced by our approach is equal to or tighter than those produced with the naive approach.
To compare to the hand-tuned approach, we select a run from \cref{fig:err_vol} that had roughly equivalent or less error in a smaller amount of time and calculate the increase in speed by our approach.
For the runs selected in \cref{fig:resultsets} and \cref{fig:resultsets2}, not every set produced by our approach is tighter than sets produced by the hand-tuned approach, but when error is calculated over all sets in the trajectory, the total error is roughly equal or smaller.
For pendulum problem S1 featured \cref{fig:resultsets}, our algorithm is 20.9\% faster than the hand-tuned approach for less error by both radius and volume metrics, and for the TORA problem T1 featured in \cref{fig:resultsets2}, our algorithm is 24\% faster for an $\approx$ equal amount of error (1.5\% less error according to the ratio metric and 1.2\% more error according to the volume metric).

In \cref{fig:car_table}, we display numerical results for the remaining problems, car problems C1 and C2.
For C1 and C2, we select runs that have less error by both radius and volume metrics and are respectively 70.8\% faster and 65.2\% faster, which translates to 4.7 minutes faster and 4.1 minutes faster.

%
%
\begin{figure}[h]
\footnotesize
    \centering
    \includegraphics[width=0.9\columnwidth,trim=0cm 0cm 0cm -.75cm,clip]{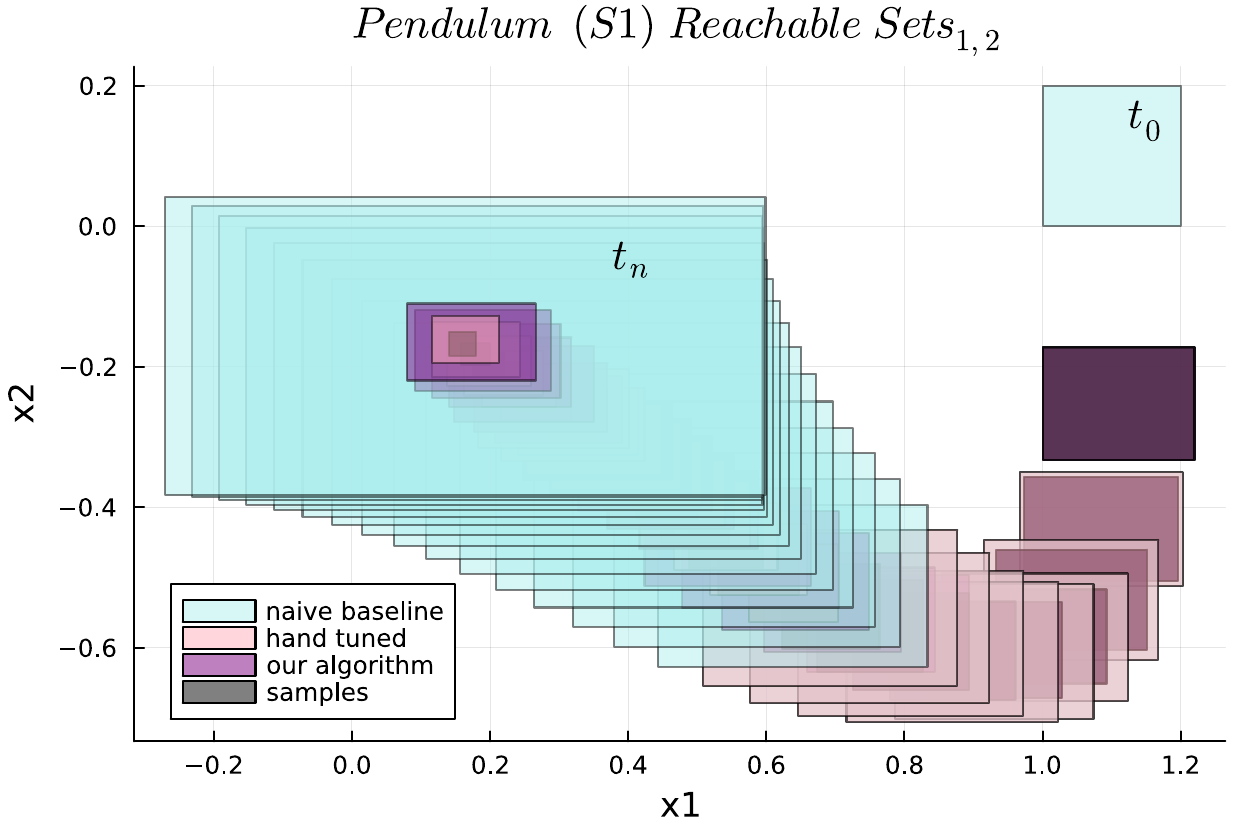}
    
\begin{tabular}{l|lll}
               & Naive & Hand-Tuned~\cite{sidrane2022overt} & Ours          \\ \hline
Error (volume) & 11.9     & 2.27       & 1.94 \\
Error (radius) & 3.79     & 1.61       & 1.59 \\
Time           & 7.84s    & 19.3s      & \textbf{15.3s}        
\end{tabular}

    \caption{\footnotesize Reachable sets for pendulum problem S1 with two layer, 25 neuron-per-layer  neural network controller show less error than the naive baseline and greater speed than the hand-tuned baseline for less error.}
    \label{fig:resultsets}
\end{figure}

\begin{figure}[h]
    \centering
    \includegraphics[width=0.9\columnwidth]{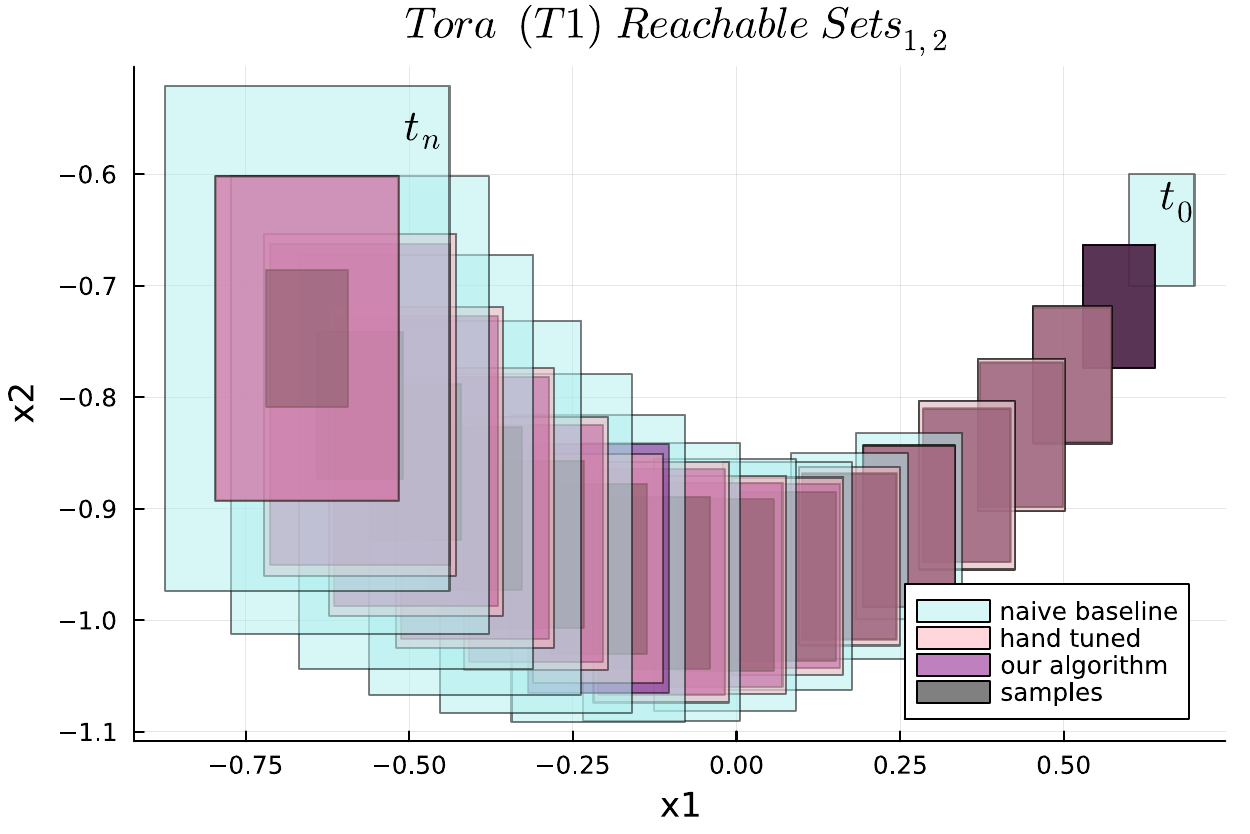}
    \footnotesize
\begin{tabular}{l|lll}
               & Naive & Hand-Tuned~\cite{sidrane2022overt} & Ours          \\ \hline
Error (volume) & 31.7     & 7.67       & 7.78 \\
Error (radius) & 1.81     & 1.41       & 1.38 \\
Time           & 6.62s    & 63.0s      & \textbf{47.8s}        
\end{tabular}

    \caption{\footnotesize Dimensions 1 and 2 of the reachable sets for TORA problem T1 with three layer, 25 neuron-per-layer neural network controller which has less error than the naive baseline and greater speed than the hand-tuned baseline for similar error.}
    \label{fig:resultsets2}
\end{figure}

\begin{figure}[h!]
    \centering
    \footnotesize
\begin{tabular}{l|lll}
\textbf{Car (C1)} & One-Step & Hand-Tuned~\cite{sidrane2022overt} & Ours          \\ \hline
Error (volume) & 37.7     & 4.90       & 3.89 \\
Error (radius) & 1.88     & 1.31       & 1.28 \\
Time           & 10.3s    & 400s     & \textbf{117s}       \\ \hline
\textbf{Car (C2)} &          &            &           \\ \hline
Error (volume)    & 37.9     & 5.29       & 4.43 \\
Error (radius)    & 1.85     & 1.30       & 1.29 \\
Time              & 11.1s    & 378s       & \textbf{131s }        
\end{tabular}
    \caption{\footnotesize Reachable sets for the Car problems C1 and C2 
    are computed significantly faster for comparable error using our algorithm as compared to the hand-tuned baseline.}
    \label{fig:car_table}
\end{figure}
%
%


\subsection{Limitations}
The approach described here has limitations related to budgeting.
A poor choice of initial solution from optimizer can lead to budget overruns as the algorithm will continue to compute if a finite solution has not yet been found.
Further, budgeting with the linear time approximation as described is not accurate for all reachability problems, leading to budget overruns and/or conservative solutions for some problems; which may be exacerbated by long horizon queries. 
As previously mentioned, our algorithm is a heuristic. 
There does not exist a guarantee that given two finite budgets $b_A > b_B$ the sets produced with budget $b_A$ will be tighter than those produced using $b_B$.
Addressing the aforementioned limitations is an area for future work.

\section{Conclusion}
In this paper we have introduced an algorithm for temporal refinement of tenuously tractable discrete time reachability problems.
Our algorithm 
makes it possible to do hybrid-symbolic reachability without hand tuning concrete and symbolic intervals.
We applied this algorithm to a difficult class of reachability problems:  nonlinear dynamical systems with neural network control policies, and demonstrated that it is able to produce reachable sets of varying error given varying computational budget.
It is also able to produce reachable sets 20-70\% faster than hand-tuned approaches for the same amount of error.
Ultimately, this algorithm 
enables more efficient reachability analysis for complex control systems.  




\bibliography{mybib}

\clearpage

\appendix
\subsection{Calculating temporal depth for Algorithm~ \ref{alg:overall}}

Algorithm~\ref{alg:calc_qsteps}, calc\_steps, calculates the next symbolic query for alg.~\ref{alg:overall}.
In line 2 of alg.~\ref{alg:calc_qsteps}, calc\_steps calculates the time index of the current step, $t_{\text{cur}}$, and then in line~3, calc\_steps makes a conservative estimate  of the amount of time needed compute one symbolic step, $\test$. 
If previously, \textit{phase = ``search''}, (line 4), and we have not stopped early (line 5), we will estimate if there is enough time budget to continue searching (line 6).

To estimate if there is enough time budget to continue searching, we must estimate the time that future symbolic queries will take.
Assuming that the time it takes to perform 1 symbolic step is roughly constant ($\test$) and that we are in the search phase, the last symbolic query was from step 0 of length $\bsteps$, and the time cost of a potential next query from time 0 to time $b_\text{steps}+1$ is $\approx t_\text{est}*(b_\text{steps}+1)$.
At any given time, the total remaining computational cost of primary\_refinement is the cost of the next search step  plus the cost of symbolic jumps to finish the time horizon in the ``jump'' phase which can be estimated as taking time $(n - (t_\text{cur}+1))*t_\text{est}$. 
As $\tcurrent = \bsteps$ during the search phase, the total time estimate for one further symbolic query from $0$ to $b_\text{steps} + 1$ plus jumps from $b_\text{steps}+1$ to $n$ is then $t_\text{est}*(b_\text{steps}+1) + (n - (\bsteps+1))*t_\text{est}= n*\test$ (see line 6). 

This estimate assumes that the time to compute $m$  symbolic steps is roughly $\test*m$  regardless of whether those steps are computed in 2 symbolic queries of size $\frac{\test*m}{2}$ or 10 symbolic queries of size  $\frac{\test*m}{10}$.
 This is a simplification  we make that is not strictly accurate -- the time a symbolic query takes may be exponential in temporal depth for some problems and timing depends on the particular input set --  but it leads to roughly accurate budgeting given that $\test$ is updated at each iteration. 

If the remaining budget $b$ does allow for continued search, $\bsteps$ is incremented (alg.~\ref{alg:calc_qsteps} line 7).
If not, we instead calculate a reasonable temporal query depth for finishing the remaining steps of the reachability problem in the jump phase (lines 8-13).
One could use jumps of size $\bsteps$ but if $n$ is not divisible or close to divisible by $\bsteps$ this could lead to a short jump at the end of the horizon and a loose final set.
 Instead, we calculate the num\_jumps needed to finish the time horizon in jumps of size $\bsteps$ (line 9-10) and instead use jumps of size $\text{ceil}(\frac{s_\text{left}}{\text{num\_jumps}})$ (line 11) which makes all the jumps closer in size.

 If the solver has stopped early (line 14), the phase is switched to ``jump'' (line 18) and a similar calculation is performed to estimate the jump size (lines 15-17)  with the difference being num\_jumps  is calculated using $\bsteps - 1$ instead of $\bsteps$ (line 16). 
 If we are already in the ``jump'' phase (line 20) we do not change the jump size of $\bsteps$ unless we stop early (line 21) in which case we decrement $\bsteps$ by 1 (line 22).
 The early stopping is used to ensure the budget  is  approximately respected given that the estimated time per symbolic step $\test$ is not an upper bound.

\begin{algorithm}
\DontPrintSemicolon
\caption{Procedure to calculate symbolic steps during primary refinement. 
\label{alg:calc_qsteps}}
\SetKwFunction{vf}{calc\_steps}
\SetKwProg{Fn}{function}{:}{}
\SetKwComment{Comment}{\#}{}
\Fn{\vf{$t_\text{start}, b_\text{steps}, b, \test$, data, $n$, phase, status}}{
$t_\text{cur} \gets t_\text{start} + b_\text{steps}$

$t_\text{est} \gets \max(t_\text{est}, \text{data}[t_\text{cur}].\text{time} / \text{data}[t_\text{cur}].\text{steps})$

\uIf{phase == ``search''}{
    \uIf{status == ``nominal''}{
        \uIf{$n*\test < b$}{
            $\bsteps \gets \bsteps + 1$\\
        }
        \Else{
        $s_\text{left} \gets n - \bsteps$\\
        num\_jumps = $\text{ceil}(\frac{s_\text{left}}{\bsteps})$\\
        $\bsteps \gets \text{ceil}(\frac{s_\text{left}}{\text{num\_jumps}})$\\
        phase $\gets$ ``jump'' 
        
        }
}\ElseIf{status == ``stopped early''}{
        $s_\text{left} \gets n - \bsteps$\\
        num\_jumps = $\text{ceil}(\frac{s_\text{left}}{\bsteps - 1})$\\
        $\bsteps \gets \text{ceil}(\frac{s_\text{left}}{\text{num\_jumps}})$\\
        phase $\gets$ ``jump'' 
        
    } 
}
\ElseIf{phase == ``jump''}{
    \If{status == ``stopped early''}{
      $\bsteps = \max(\bsteps - 1, 1)$  
    }
}
    return $b_\text{steps}$, phase
\;}
\end{algorithm}

\end{document}